\documentclass[graybox]{svmult}
\usepackage[utf8]{inputenc}

\title{HLRS2023}
\author{xu  chu }

\usepackage{float}
\usepackage{mathptmx}       
\usepackage{helvet}         
\usepackage{courier}        
\usepackage{type1cm}        

\usepackage{makeidx}         
\usepackage{graphicx}        
\usepackage{subfigure}
\usepackage{caption}
\usepackage{multicol}        
\usepackage[bottom]{footmisc}
\usepackage{multirow}
\usepackage{latexsym}
\usepackage{epstopdf}
\usepackage[square]{natbib}
\usepackage{authblk}
\usepackage{amsmath,bm}
\usepackage{amsmath}
\usepackage{amssymb}

\makeindex             

\begin{document}

\title*{Modeling of Coupled Turbulent Channel-Porous Media Flow through a Deep Autoencoder-Echo State Network Framework}
\titlerunning{Turbulence modulation and energy transfer above porous media}


\author{Xu Chu \inst{1} \and Sandeep Pandey \inst{2} \and Yanchao Liu \inst{3} \and Bernhard Weigand \inst{3}}
\institute{\inst{1} Cluster of Excellence SimTech, University of Stuttgart, Pfaffenwaldring 5a, 70569 Stuttgart, Germany,  \email{xu.chu@simtech.uni-stuttgart.de} \and %
                      \inst{2} Institute of Thermodynamics and Fluid Mechanics, Technische Universit{\"a}t Ilmenau, D-98684 Ilmenau, Germany \and %
                    \inst{3} Institute of Aerospace Thermodynamics (ITLR), Pfaffenwaldring 31, 70569 Stuttgart, Germany  
                      }
\authorrunning{X.Chu, S. Pandey, Y. Liu, B.Weigand}                      

\maketitle
\vspace{-2cm}
\abstract {
In this study, we propose a novel approach, namely the combined Convolutional Deep Autoencoder–Echo State Network (CDAE-ESN) model, for the analysis and forecasting of dynamics and low-order statistics in coupled turbulent channel-porous media flows. Such systems find wide applications in industrial settings, including transpiration cooling and smart interface engineering. However, the complex geometry of coupled flow systems presents additional challenges for purely data-driven models.
Our results demonstrate that the integration of deep autoencoder and echo state network techniques enables effective modeling and prediction of dominant flow behaviors, particularly within the porous domain exhibiting laminar regimes. To enhance the model's applicability across a broader range of data domains, we further employ fine-tuning on a dataset encompassing varying porosities. The achieved average statistics exhibit a reasonable agreement, underscoring the efficacy of our proposed approach.
 }

\,

\section{Introduction}
\label{sec:1}
Turbulent flow over permeable interfaces is ubiquitous in nature and engineering applications, from sediment transport to the transpiration cooling in gas turbines. Porous media are generally featured with complex topology in space \citep{Chu.2018}. A wide range of properties has been identified to have a significant effect on the mass and momentum transport across the porous medium-free flow interface. However, understanding and optimizing macro/micro-scale characteristics of porous media remains a challenge. This is due to both the difficulty of experimental measurements at the pore scale \cite{Terzis.2019} as well as the high costs of numerical simulations to resolve the full spectrum of scales.

Early studies of turbulent flows over permeable beds have addressed the effect of varying permeability on surface flows. The turbulent surface flow is similar to that of a canonical boundary layer when the interfacial permeability is low, as the near-wall structures are less disrupted. However, as the permeability increases, large-scale vortical structures emerge in the surface flow. This is attributed to Kelvin-Helmholtz (KH) type instabilities from inflection points of the mean velocity profile. More recently, the application of porous media for drag reduction inspired a series of in-depth studies on anisotropic permeability. 

Direct Numerical Simulation (DNS) exhibits an edge of observing and analyzing turbulent physics in a confined small space, not only for the canonical cases as channel flows and pipe flows \cite{Chu.2016, Pandey.2017b}. Suga et al. \cite{Suga.2020} investigated the influence of the anisotropic permeability tensor of porous media at a higher permeable regime. It was found that streamwise and spanwise permeabilities enhance turbulence whilst vertical permeability itself does not. In particular, the enhancement of turbulence is remarkable over porous walls with streamwise permeability, as it allows the development of streamwise large-scale perturbations induced by the Kelvin–Helmholtz instability. However, DNS is still an expensive solution when one extend it for the parametric study often desired in the industrial context. It is still common to use analytical models or advanced turbulence models for this purpose which provide a reasonable accuracy in established applications. However, these models can diverge and model calibration is a time-consuming process.

Machine learning (ML) is on surge where it has successfully tackled many complex tasks ranging from self-driving cars, drug discovery, weather prediction, and more recently state-of-the-art large language models such as GPT for natural language processing. Machine learning has been democratized heavily in the last decade and some of the reasons include vast availability of open-source libraries, active community, huge volume of incoming data from physical as well as numerical experiments and an ease in access to high-performance computing resources via cloud. In recent years, ML has shown potential in the modelling of various fluid flow use-cases \citep{pandey2022direct, fang2019neural, chu2018computationally, pandey2020perspective}. Typically, big data ML problems rely upon some kind of feature extraction process e.g. conventional feature selection process, long-standing proper orthogonal decomposition (POD), relatively new dynamic mode decomposition (DMD) \cite{kutz2016dynamic} or ML based methods \citep{storcheus2015survey}. Feature extractor transform the input raw data into information preserving feature vector with a primary goal of reducing the dimensionality of the data. Several attempts have been made in the past few years to combine POD or DMD with machine learning \citep{pandey2020reservoir}. However, these methods have certain limitation especially when the flow advances in the turbulent flow regimes. Therefore, an end-to-end machine learning system is being suggested where POD and DMD is replaced by a non-linear DNN \citep{pandey2020reservoir, nakamura2021convolutional}. A common architecture is deep convolution autoencoder (DAE) to extract the features for further processing and decision making. Turbulent flow problems often involve tempo-spatial data, therefore, extracted features could be used to train a downstream network to build a data-driven dynamic model. A common choice is a recurrent neural network (RNN) which preserves the sequential relationship in a time series. A combination of such approaches has shown exemplary results in terms of data compression and modeling.

In this work, we make an attempt to model complex flow behaviors in a porous medium with the help of a data-driven approach. The flow is in a turbulent regime and due to the presence of porosity, geometry is resolved to a finer scale. This setup makes the end-to-end ML task challenging and equally interesting. The goal is to combine the deep learning based feature extraction step followed by a recurrent neural network to model temporal evolution. Echo state network was chosen as the RNN due to its proven efficiency in modeling multi-dimensional tempo-spatial data. 
The trained models are restricted to the domain of training and work poorly beyond the trained data domain which could be a specific parameter such as Reynolds number. A commonly used strategy called transfer learning is used, where the objective is to transfer the gained knowledge from the training of source task to related target task with a limited dataset \cite{zhuang2020comprehensive}. 
Fine tuning, which can be considered as a part of transfer learning, is one methodology where an already trained and optimized model tuned with another similar set of data for a similar task. This could decrease the training time significantly in addition to neural architecture search and training data requirements. Fine-tuning can be performed on the entire network level where we allowed to retrain all the layers while initializing the parameters with the pretrained model. Alternatively, several layers can be freeze and retraining is allowed on only few layers. The former method is suitable when network size is very large and data is limited. While first approach could lead to the overfitting \cite{yosinski2014transferable}. Therefore, we also present the result from fine-tuning where trained model from a given data domain was transferred to a different data domain.

\section{Synthetic data generation}
\label{synthetic_data_generation}

\begin{figure}[ht!]
    \centering
	 \includegraphics[width=6cm, keepaspectratio]{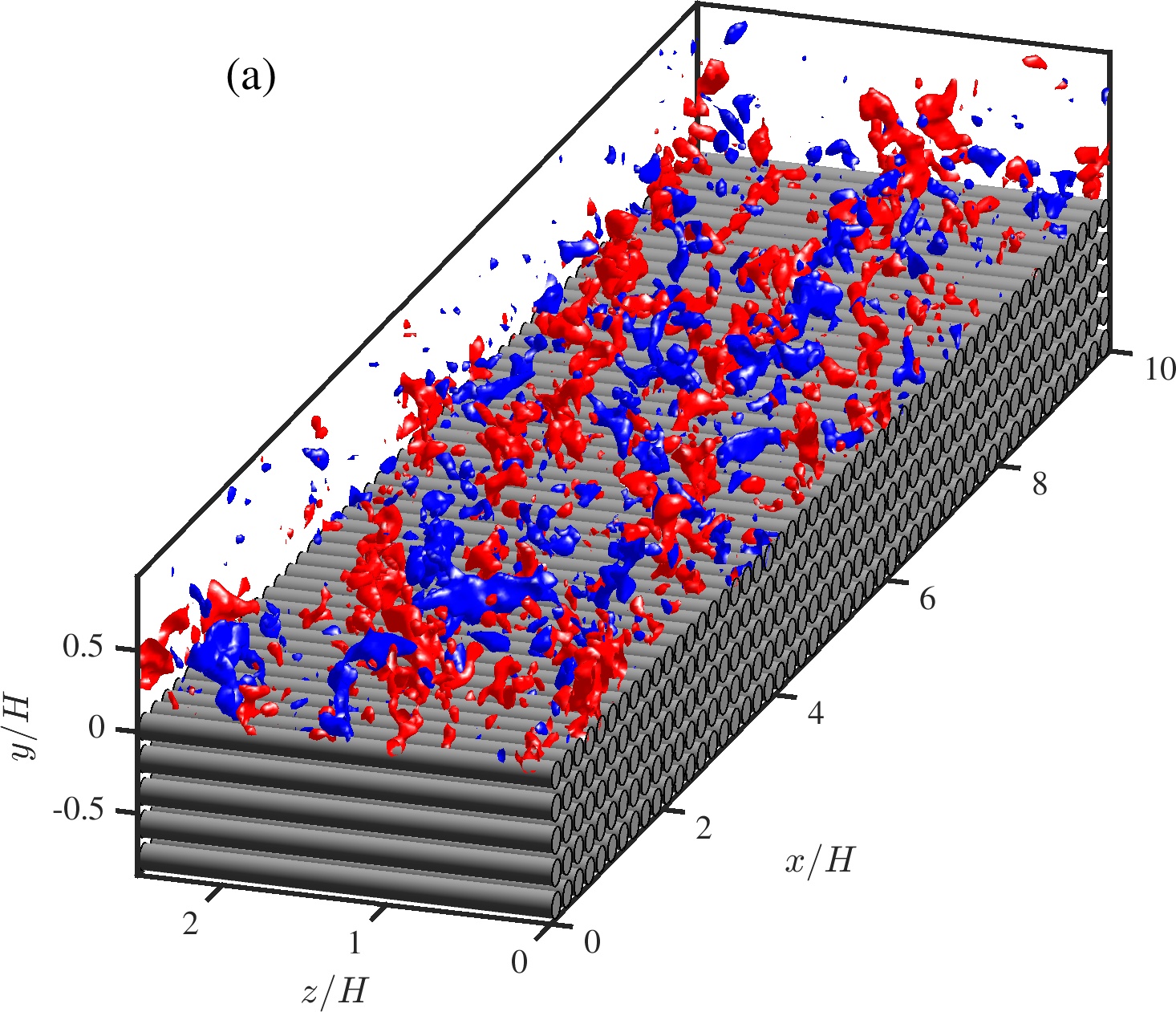}    \\
	\caption{Configuration of the computational domain (C05), the permeability in three directions can be found in Table \ref{tab:1}. The blue and red isosurfaces show the wall-normal fluctuation $v'/u^p_\tau$ at level $-0.4$ and $0.4$, respectively. }
	\label{fig:setup}
\end{figure}

In our DNS, the three-dimensional incompressible Navier$-$Stokes equations are solved in non-dimensional form, 

\begin{equation}
\frac{\partial u_j}{\partial x_j}=0
\label{eq:1}
\end{equation}
\begin{equation}
\frac{\partial u_i}{\partial t}+\frac{\partial u_i u_j}{\partial x_j}=-\frac{\partial p}{\partial x_i}+\frac{1}{Re_D}\frac{\partial^2 u_j} {\partial x_i \partial x_j}+\Pi \delta_{i1}
\label{eq:2}
\end{equation}

where $\Pi$ is a constant pressure gradient in the mean-flow
direction.  The governing equations are normalized using the half-width of the whole simulation domain $H$(figure \ref{fig:setup}) and the averaged bulk velocity $U_b$ of the channel region ($y/H=[0,1]$). Hereafter, the velocity components in the streamwise $x$, wall-normal
$y$, and spanwise $z$ directions are denoted as $u$, $v$, and $w$,
respectively.  The domain size ($L_x/H \times L_y/H \times L_z/H$) is $10
\times 2 \times 0.8\pi$ in all cases. The lower half ($y/H=[-1,0]$) contains the porous
media, and the upper half ($y/H=[0,1]$) is the free channel flow. The
  porous layer consists of 50 cylinder elements along the
  streamwise direction and 5 rows in the wall-normal
  direction, as illustrated in figure \ref{fig:setup}. The distance
$D$ between two nearby cylinders is fixed at $D/H=0.2$. No-slip boundary conditions are applied to the cylinders, the upper wall, and the lower wall. Periodic boundary conditions are used in both
streamwise and spanwise directions.


The spectral/$hp$ element solver Nektar++ is used to solve equations (\ref{eq:1}, \ref{eq:2}) \citep{Cantwell.2011, Chu.2019, Chu.2020,Pandey.2020, Chu.2021, Chu.2021b, Wang2021assessment, Wang.2022}. The geometry in the $x$-$y$ plane is discretized into quadrilateral elements with local refinement near the cylinders (see figure \ref{fig:setup}). Local element expansions are applied based on the modified Legendre basis \citep{karniadakis2005spectral}. We used flexible polynomial orders across the wall normal range in a continuous Galerkin projection. The polynomial order in the free flow region $y/H=[0.2, 1]$ is $P=6$-$7$. The near wall region and the top two layers of cylinders $y/H=[-0.4, 0.2]$ are enhanced with a higher order of $P=8$-$9$. A lower order of $P=5$ is selected in the deeper positions of the cylinder array ($y/H=[-1, -0.4]$). The spanwise direction is extended with a Fourier spectral method. The 2/3 rule is used to avoid aliasing errors. The time-stepping is performed with a second-order mixed implicit-explicit (IMEX) scheme. The time step is fixed at $\Delta T/(H/U_b)=5\times10^{-4}$.

\begin{table}
\centering
\begin{tabular}{ccccccccccc}

Case&$\varphi$  & 
$Re_{\tau}^{p}$&$Re_{\tau}^{s}$&$\sqrt{K_{xx}}^{p+}$,$\sqrt{K_{yy}}^{p+}$ 
 &$\sqrt{K_{zz}}^{p+}$ & $C_{xx}^{{p+}},C_{yy}^{{p+}}$ & $r_c^{{p+}}$& $C_f^p$ & $C_f^s$\\

C05  &0.5 & 336  & 180  &4.55&8.86&0.37&42&0.0112&0.0084\\
C06  & 0.6 & 464   & 190  &9.34&15.23&0.58&48&0.0149 &0.0085 \\

\end{tabular}
\caption{Simulation parameters. The porosity of the porous medium
  region is $\varphi$. The friction Reynolds numbers are $Re_{\tau}^{p}$ and
  $Re_{\tau}^{s}$ for the porous and impermeable top walls respectively. $\sqrt{K_{\alpha\alpha}}^{p+}$ and $C_{\alpha\alpha}^{{p+}}$ are the diagonal components of the permeability tensor and Forchheimer coefficient, respectively, in the direction of $\alpha$ ($\alpha\in\{x, y,z \}$), which are normalized by wall units. $r_c^{{p+}}$ is the radius of the cylinders.}
\label{tab:1}
\end{table}

Four DNS cases are performed with varying porosity $\varphi=0.5$ and 0.6, which is
defined as the ratio of the void volume to the total volume of the
porous structure. The parameters of the simulated cases are listed in table \ref{tab:1}, where the cases are named after their respective porosity. The superscripts $(\cdot)^p$ and
$(\cdot)^{s}$ represent permeable wall and smooth wall side variables, respectively. Variables with superscript $^+$ are scaled by friction velocities $u_\tau$ of their respective side and viscosity $\nu$. 

For an arbitrary variable $\phi$, we denote the time averaged value as $\overline{\cdot}$, i.e., $\overline \phi=1/T\int_{0}^T\phi \mathrm{d}t$, and spatial averaged value in $x$-$z$ plane as $\langle \cdot \rangle$, i.e., $\langle \phi \rangle=1/A_f \int_{A_f}\phi \mathrm{d}A$, $A_f$ being the fluid  area.  The instantaneous turbulent fluctuation is $\phi'=\phi-\overline{\phi}$, and form-induced fluctuation is  $\widetilde{\overline{\phi}}=\overline{\phi}-\langle \overline{\phi} \rangle$. For the permeable wall side, the total shear stress at the interface can be derived as:

\begin{equation}
\tau_{w}^p= \left ( \mu \frac{\partial \langle \overline{u} \rangle}{\partial y} - \rho\langle\overline{ u^{\prime}v^{\prime}}\rangle-\rho \langle{\overline{u}}\rangle \langle{\overline{v}}\rangle -\rho\langle{ \tilde{\bar{u}}   \tilde{\bar{v}}}\rangle\right )_{y=0}.
\label{eq:tau}
\end{equation}
For the smooth wall side, 
\begin{equation}
\tau_{w}^s= \left ( \mu \frac{\partial \langle \overline{u} \rangle}{\partial y}\right )_{y=H}, 
\label{eq:tau2}
\end{equation}
The friction velocity can then be calculated from total shear stress at both walls $u_\tau=\sqrt{\tau_w/\rho}$. The friction coefficient is defined as $C_f=\tau_{w}/(\frac{1}{2}\rho U_b^2)$, which is listed in table \ref{tab:1}. 

Note that the distance between cylinders is fixed, and the porosity is changed by varying the radius of the cylinders. The normalized cylinder radius is in the range $r_c^{p+}=r_c u_\tau^{p}/\nu=42$-$48$ for all the cases tested (see table \ref{tab:1}), such that the effect of surface roughness is assumed to be at a similar level. For all the cases, the Reynolds number of the top wall boundary layer is set to be $Re_\tau^{s}=\delta^{s}u_\tau^{s}/\nu\approx180$ ($\delta$ is the distance between the position of maximum streamwise velocity and the wall). In this manner, changes in the top wall boundary layer are minimized. On the top smooth wall side, the streamwise cell size ranges from $4.1\le\Delta
x^{s+}\le6.3$ and the spanwise cell size is below $\Delta
z^{s+}=5.4$. On the porous media side, $\Delta z^{p+}$ is below 8.4, whereas $\Delta x^{p+}$ and $\Delta y^{p+}$ are enhanced by polynomial refinement of local mesh. The total number of grid points ranges from 88$\times10^6$ (C05) to 595$\times10^6$ (C08). Each cylinder in the porous domain is resolved with 80 to 120 grids along the perimeter.
The spatial resolution of current work is close to those of previous DNS researches. For example, \cite{Shen.2020} conducted DNS of a turbulent flow over sediment beds using the immersed boundary method, where the diameter is discretized into 36-50 points. The mesh of \cite{Karra.2022} has 26 grid points along the diameter of each grain.

By performing parameter tests on porous media and fitting the Darcy-Forchheimer equation, the permeability tensor $K$ and Forchheimer coefficient $C$ may be obtained, and the details of the computation can be referred to \cite{Wang.2021}. The intensity profiles of $u'$, $v'$, and $w'$ are shown in figure \ref{fig:setup}(b-d), respectively. The turbulent intensity grows with porosity for both the channel and the porous medium region. In the current study, we focus on the up and down-welling motions at the upper surface of the porous medium, which is represented by the fluctuation $v'$. 

\section{Data-driven modelling}
\label{ml_framework}

\begin{figure}[ht!]
\centering
\includegraphics[width=13cm, keepaspectratio]{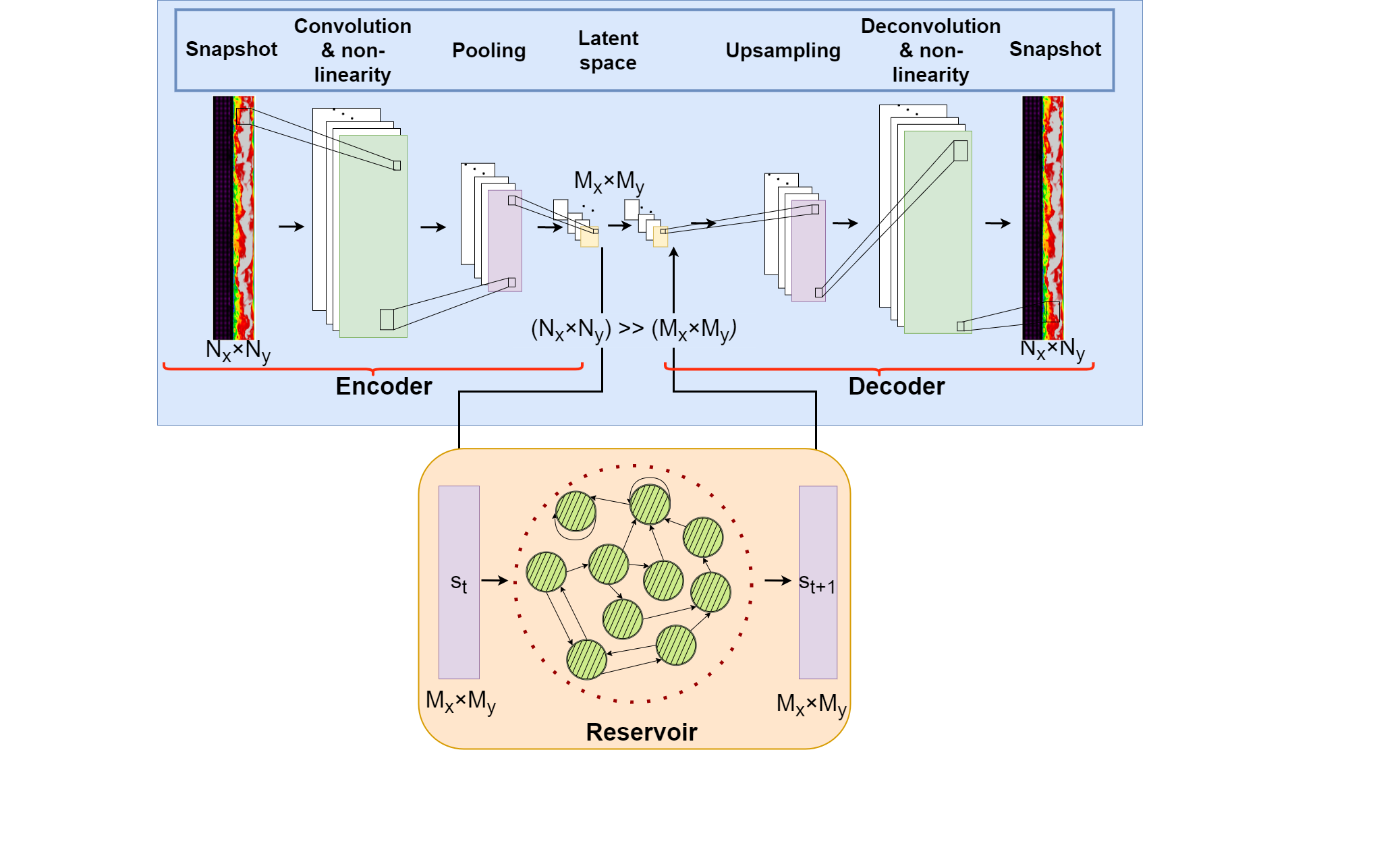}
\caption{\label{auto_results} Two-step machine learning architecture for end-to-end ROM model for porous media. In the first step, DAE is trained as feature extractor cum decoder while the second step trains the extracted feature in autoregressive manner.}
\label{Fig-2}
\end{figure}

Data driven modelling (DDM) is an approach to harness the available data about
a system and establish a connection between the system state variables (input,
internal and output variables) without explicit knowledge of the physical behaviour \cite{solomatine2008data}. In this work, we have the synthetic data derived from the DNS while the end model should be based upon this data and should able to provide an model to predict temporal evoliution of a state variable. Figure \ref{Fig-2} depicts our end-to-end 2-step ML framework. To have an extendable framework, we chose to first extract the features with the help of a deep encoder and feed the extracted features to a second level recurrent neural network (RNN). This architecture is inspired from our earlier work \cite{pandey2022direct}. 

\subsection{Feature extraction with autoencoder}
\label{dae}

The deep convolution autoencoder (DAE) is an implementation of an unsupervised machine learning paradigm, where the training of the model does not rely upon the labeled data. 
The objective of learning is to extract efficient representations or features in a low-dimension hyper-plane that can reproduce the data with minimal reconstruction error. 
That is to say, an autoencoder encodes input data $\bm{I}$ into $\bm{c}=\textnormal{encode}(\bm{I})$ with $\bm{c} \in \mathbb{R}^l$. And it decodes $\bm{c}$ back into $\hat{\bm{I}}=\textnormal{decode}(\bm{c})$ subsequently.
The reconstruction error is calculated with $\hat{\bm{I}}-\bm{I}$.
Here, 2-dimensional data were obtained from the DNS as described in Sec. \ref{synthetic_data_generation}. Thus autoencoder consists of several convolution layers resulting in a deep architecture. Each convolution layer in the network acts as a local feature extractor and shares the weights. The convolution layer is followed by a pooling layer, max pooling in this case, which reduces the dimension and focuses on the non-repeating key features. It gives a unique property of translation invariance to the network. At the end of the encoder network, there is a bottleneck layer that has the latent space or extracted features. These features are then passed to a decoder section which will attempt to recreate the original DNS data from the features. The training starts with random initialization of all the weights in the network and learns them as per the loss function via backpropagation.

\subsection{Echo State Network (ESN)}
\label{dae}
ESN is closely related to reservoir computing (RC) and it shines out with sequential data where temporal relationship is present. In a typical ESN setup, input data is fed to a input layer which is connected to a big sparse reservoir which is finally connected to an output layer. ESN differs from typical RNN in terms of not requiring a back-propagation, therefore, it is inherently fast in training. More details can be found in our earlier work \cite{pandey2022direct}. In this work, ESN acts as a downstream network which learns to predict the temporal evolution of the features obtained by the encoder section of DAE in an auto-regressive manner. The predicted values can be fed back to the decoder section of DAE to reproduce the feature of interest. Therefore, the during the deployment, we are only interested in the trained ESN and decoder while encoder can be thrown away. 

\section{Results and Discussion}
\label{dae}

\subsection{Data driver flow model}
\label{dae}
As mentioned earlier, we first trained a DAE network to extract the feature in a lower-dimensions. For the same, input and output layer has the same dataset i.e. the snapshots from the DNS consist of $u$-velocity component with a dimension of $1501 \times 109 \times 1$. The neural network has multiple convolution layers combined with a max pooling layer while the decoder section has a convolution layer and a up-sampling layer, giving a converging-diverging shape. The features or latent modes are obtained at the end of the encoder section and it has a shape of $24 \times 2 \times 34$, thereby giving a dimensionality reduction of 99\%. This could increase further on the expense of reconstruction loss. Table \ref{tab:cae} illustrates the various layers and their shape. After fixing the DAE architecture, various hyperparameters were obtained by using Bayesian optimization. This particular approach is part of the AutoML paradigm \cite{he2021automl}, and it assists in achieving an optimized set of parameters in a relatively less number of iterations when compared to random or grid search. It is worth to mention that the data were scaled between 0-1 before the training.

\begin{table*}[htb]
 \begin{tabular}{l c c l c} 
\hline\hline
\multicolumn{2}{c}{Encoder} & $\quad\quad\quad\quad\quad$ & \multicolumn{2}{c}{Decoder} \\
Layer & Output size & & Layer & Output size \cr 
\hline
 Encoder input & $1501\times 109 \times 1$ & & Decoder input & $24\times 2 \times 34$ \cr 
 2D Conv-E1 & $1501\times 109 \times 256$ & &2D Conv-D1 & $24\times 2 \times 34$ \cr 
 Max Pool-E1 & $751\times 55 \times 256$ & & 2D Upsamp-D1 & $48\times 4 \times 34$ \cr 
 2D Conv-E2 & $751\times 55 \times 256$ & & 2D Conv-D2 & $48\times 4 \times 32$ \cr 
 Max Pool-E2 & $376\times 28 \times 256$ & & 2D Upsamp-D2 & $96\times 8 \times 32$ \cr 
2D Conv-E3 & $376\times 28 \times 128$ & & 2D Conv-D3 & $96\times 8 \times 64$ \cr 
Max Pool-E3 & $188\times 14 \times 128$ & & 2D Upsamp-D3 & $192\times 16 \times 64$ \cr 
 2D Conv-E4 & $188\times 14 \times 64$ & & 2D Conv-D4 & $192\times 16 \times 128$ \cr 
 Max Pool-E4 & $94\times 7 \times 64$ & & 2D Upsamp-D4 & $384\times 32 \times 128$ \cr 
  2D Conv-E5 & $94\times 7 \times 32$ & & 2D Conv-D5 & $384\times 32 \times 256$ \cr 
 Max Pool-E5& $47\times 4\times 32$ & & 2D Upsamp-D5 & $768\times 64 \times 256$ \cr 
  2D Conv-E6 & $47\times 4 \times 34$ & & 2D Conv-D6 & $768\times 64 \times 256$ \cr 
 Max Pool-E6 & $24\times 2\times 34$ & & 2D Upsamp-D6 & $1536\times 128 \times 256$ \cr 
  &  &  & Output with 2D Conv & $1536\times 128 \times 1$ \cr 
  & &  & Output with Cropping & $1501\times 109 \times 1$ \cr 
\hline\hline
\end{tabular}
\caption{DAE architecture used at first level for feature extraction. The table summarizes the encoder and decoder architectures (Conv = convolution, Max Pool = max pooling, Upsamp = upsampling). Symbol E2 denotes for example encoder hidden layer No. 2.}
\label{tab:cae}
\end{table*}

\begin{figure}[ht!]
\centering
\includegraphics[width=8cm, keepaspectratio]{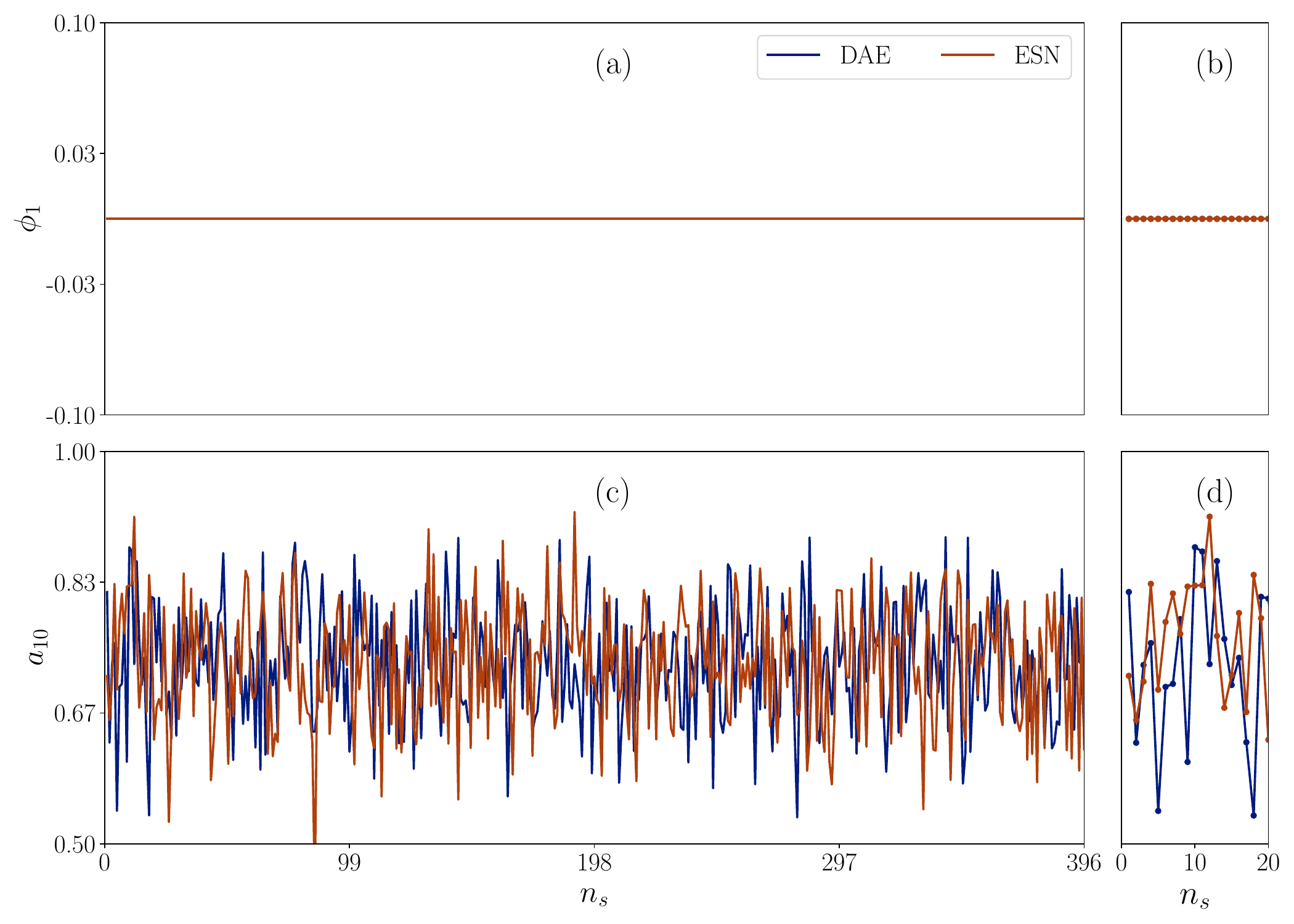}
\caption{\label{auto_results} A comparison of extracted features from DAE which acts as ground truth for the training, and prediction from ESN for 2 of the modes.}
\label{Fig-3}
\end{figure}

\begin{figure}[ht!]
\centering
\includegraphics[width=8cm, keepaspectratio]{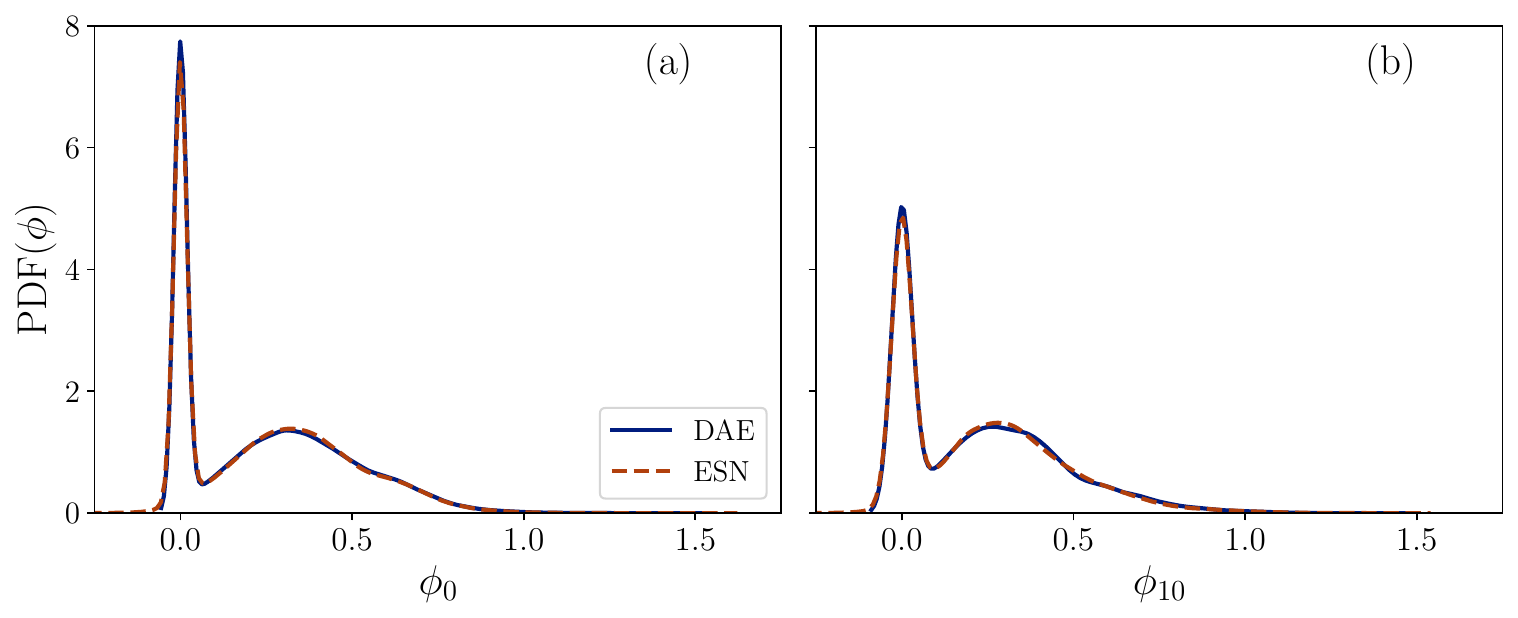}
\caption{\label{auto_results} Comparison between DAE and ESN modes in terms of pdf for 2 of the modes}
\label{Fig-4}
\end{figure}

Once DAE is trained, we extract the features for the entire dataset using its encoder part. These encoded data which has 99\% reduced dimension were further used to train the second level of RNN i.e. ESN. Unlike DAE, training an ESN is relatively straightforward and computationally cheaper because of straightforward design choices. The optimized network of ESN has 3726 reservoirs, 0.93 as spectral radius, 0.94 as leaking rate and 0.16 as reservoir density. For the best results, the look back history period is 4 timesteps i.e., model requires 4 timesteps data as an initialization then the system can predict auto-regressively. As a first step of validation, we visualize the low-dimensional features from the DAE along with the generated data from ESN on the blind validation set. We call the method generative because the model works auto-repressively once we provide the initialize data. Figure \ref{Fig-3} shows such time-series for 2 of the modes out of 1632. It is impressive that model maintain a value of 0 without any minor fluctuations for $\phi_1$ while series fluctuates with the right phase and amplitude for $\phi_{10}$. Figure \ref{Fig-4} further illustrates the probability density function (PDF) of these 2 modes for the DAE and ESN. It reaffirms the generating capability of ESN where it is able to match the overall distribution with the ground trouth. 

\begin{figure}[ht!]
\centering
\includegraphics[width=12cm, keepaspectratio]{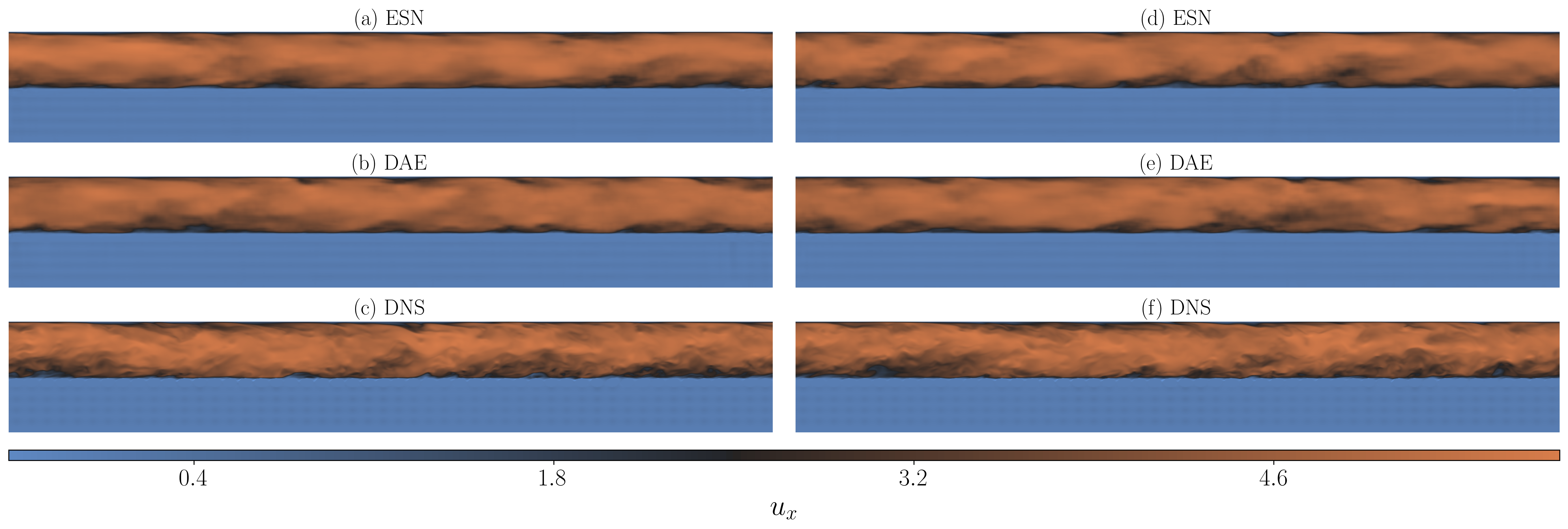}
\caption{\label{auto_results} Contours of $u_x$ by (a,d) ESN (final end to end model), (b,e) DAE (feature extractor) and (c,f) DNS (GT) at 2 different timestamps.}
\label{Fig-5}
\end{figure}

\begin{figure}[h!]
\centering
\includegraphics[width=8cm,keepaspectratio]{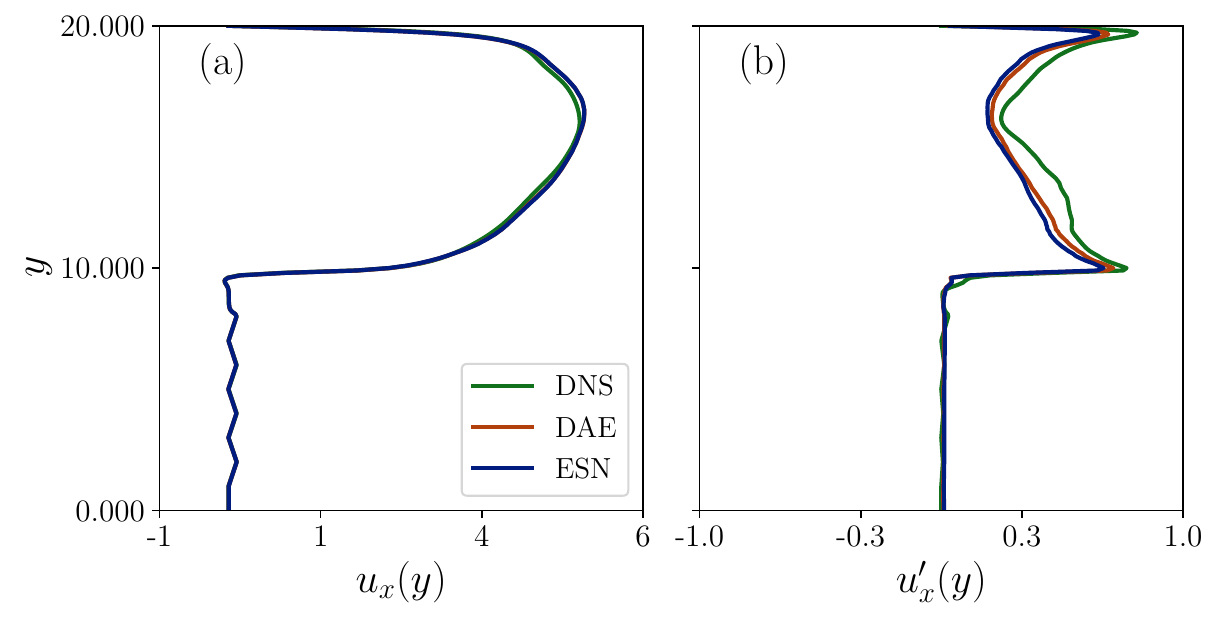}
\caption{\label{auto_results} Comparison among ground truth (DNS), reduced order model (DAE) and temporal model for DAE modes using ESN. (a): Average fields (b) Fluctuations}
\label{Fig-6}
\end{figure}

As a next step of validation is the comparison in high-dimensional space where flow physics need to be captured. Figure \ref{Fig-5} displays contour plots of the original DNS data in panels (c) and (f), reconstructed data from the features space of the DAE in panels (b) and (e), and the decoded value from the prediction of ESN is shown in panels (a) and (d). One can observe a good qualitative agreement in terms of capturing the flow fields. Both regions i.e. main channel flow and porous flow can be predicted. Both DAE and ESN were trained from the DNS, and are not able to fully capture the strong turbulent field in the channel. While the flow in the porous media section is much better because it falls within the laminar region mostly. The obvious reason is the limited capability of DAE to capture the entire flow field because it discards 99\% of input data and usage only 1\% of the features, similar to conventional reduced-order modelling approach such as POD. The derived model with ESN shows an excellent result while matching field compared to its GT i.e. DAE. It suggests that the entire model can further be improved if a state-of-the-art feature extractor such as transformer model can be used instead of DAE. It is worth to mention DAE surpasses conventional as well as some advance feature extractors e.g., variational autoencoder for this use-case, which was observed in our experimentation. Often, average and fluctuations are critical values for the analysis in industrial applications. Therefore, we further validate the results with mean and fluctuations profile over the cross-section of the channel. Figure \ref{Fig-6} shows these profiles and a mean velocity profile overlap with the DNS for both DAE and ESN. The fluctuation of velocity is slightly deviated in the main channel while they remain zero in the porous section due to dominating laminar flow. The distinction made by the model is also great where it can predict the interface where flow is transiting from laminar to turbulent.

\subsection{Fine tuning}
\label{ft}
The trained model has shown a good agreement in terms of quantitative as well as qualitative characteristics of the flow. However, the trained model is limited to a given data domain that was used for the training which means that the model will not predict accurate results as we use different conditions. This is one of the biggest downsides of such a data-driven model. There are several approaches that one can use to increase the applicability of such a model to a wider data domain. In this work, we assess the fine-tuning which is widely used in the machine-learning community. Fine-tuning makes the use of existing trained model and tune it with the newer limited data. As mentioned earlier, DAE is a computationally expensive component of this workflow, therefore, we further assess DAE with the fine-tuning. We tested the DAE model with a different set of data. Here, we used case C06 which has a higher level of porosity compared therefore a higher Reynolds number as mentioned in Table \ref{tab:1}. 

\begin{table*}[htb]
\begin{tabular}{c c c c} 
 \hline
 Case Name & Description & Training time & Mean Square Error \\ [0.5ex] 
 \hline\hline
 DNS & Simulation results which acts as GT for case C06 & - & - \\ 
 \hline
 DAE90 & Model trained from scratch similar to case C05 & O(10$^3$) & 0.025 \\
 \hline
 DAE91 & Direct prediction using model from case C05 without any training & 0 & 0.081 \\
 \hline
DAE92 & Model from case C05 retrained with all data points for case C06 & O(10$^2$) & 0.030 \\
 \hline
 DAE93 & Model from case C05 retrained with 10\% data points for case C06 & O(10$^1$) & 0.049 \\
 \hline
\end{tabular}
\caption{Various cases considered for fine tuning.}
\label{tab:fine_tune}
\end{table*}

\begin{figure}[ht!]
\centering
\includegraphics[width=10cm, keepaspectratio]{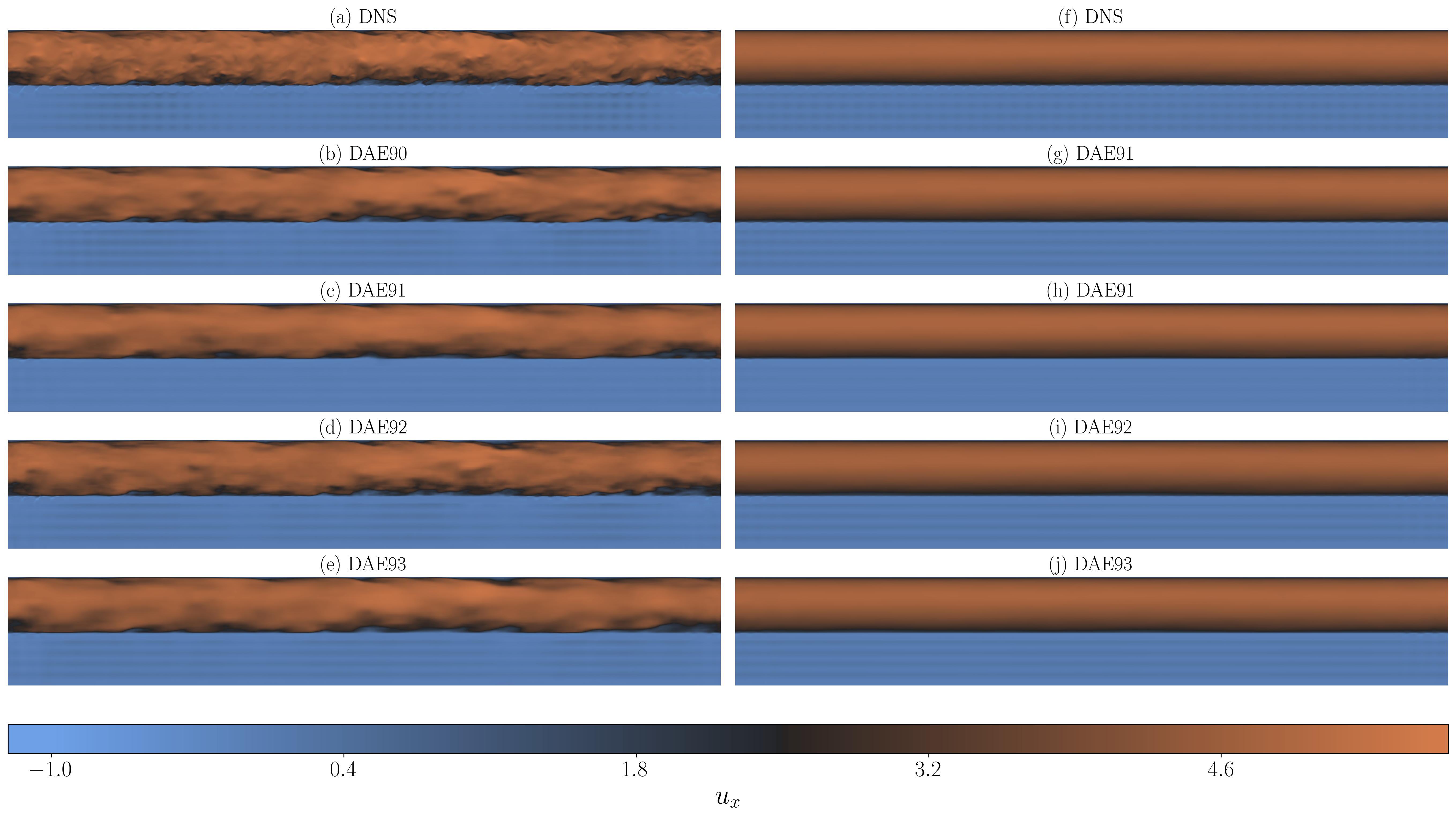}
\caption{\label{auto_results} Contours of (a, b, c, d): Instantaneous velocity, (f, g, h, i): Temporal average velocity. Labels are as per Tab. \ref{tab:fine_tune}}
\label{Fig-7}
\end{figure}

\begin{figure}[h!]
\centering
\includegraphics[width=8cm,keepaspectratio]{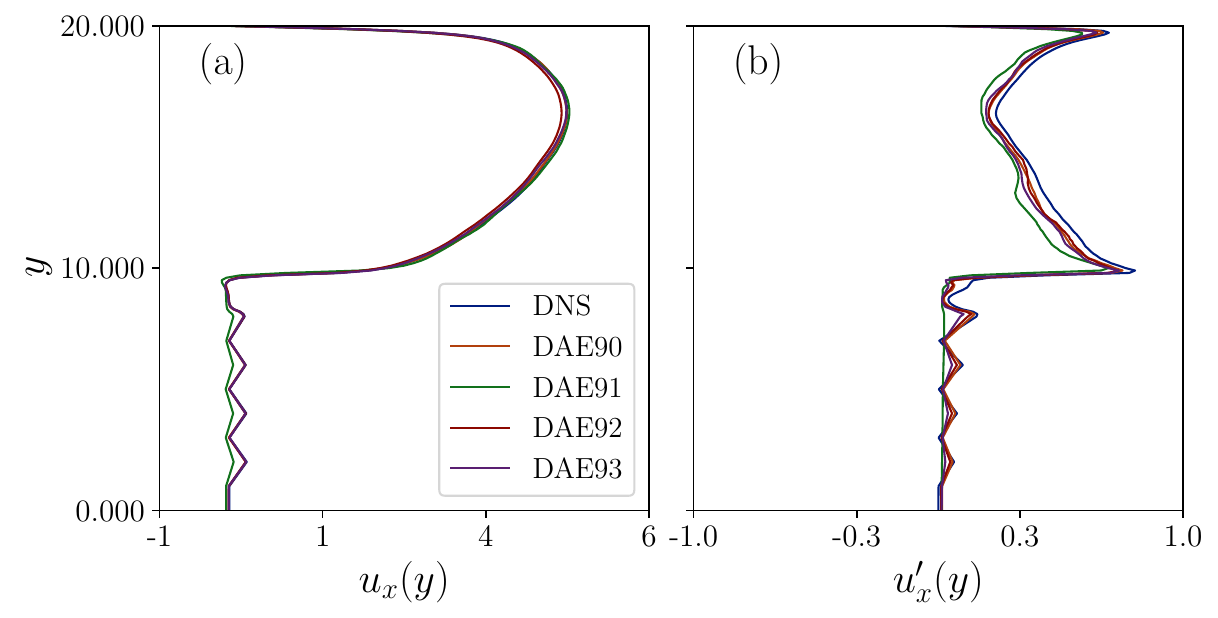}
\caption{\label{auto_results} Comparison among ground truth (DNS), reduced order model (DAE) and temporal model for DAE modes using ESN. (a): Average fields (b) Fluctuations}
\label{Fig-8}
\end{figure}

Table \ref{tab:fine_tune} shows the cases used for the fine tuning study and corresponding results are shown in Figures \ref{Fig-7} and \ref{Fig-8}. For the averaged streamwises velocity profile and its fluctuation, the reduced-order model shows comparable results as in its original validation, which proves the validity of the methodology of the fine-tuning.

\section{Computational performance}
\label{sec:5}

The supercomputing systems used for the DNS was {\it HAWK} located at the High Performance Computer Center Stuttgart (HLRS). The new flagship machine {\it HAWK}, based on the Hewlett Packard Enterprise platform running AMD EPYC 7742 processor code named {\it Rome}, has a theoretical peak performance of 26 petaFLOPs, and consists of a 5,632-node cluster. One AMD EPYC 7742 CPU consists of 64 cores, which leads to 128 cores on a single node sharing 256 GB memory on board. This means that a total of 720,896 cores are available on the HAWK system.

\begin{figure}
\centering
\includegraphics[width=4in]{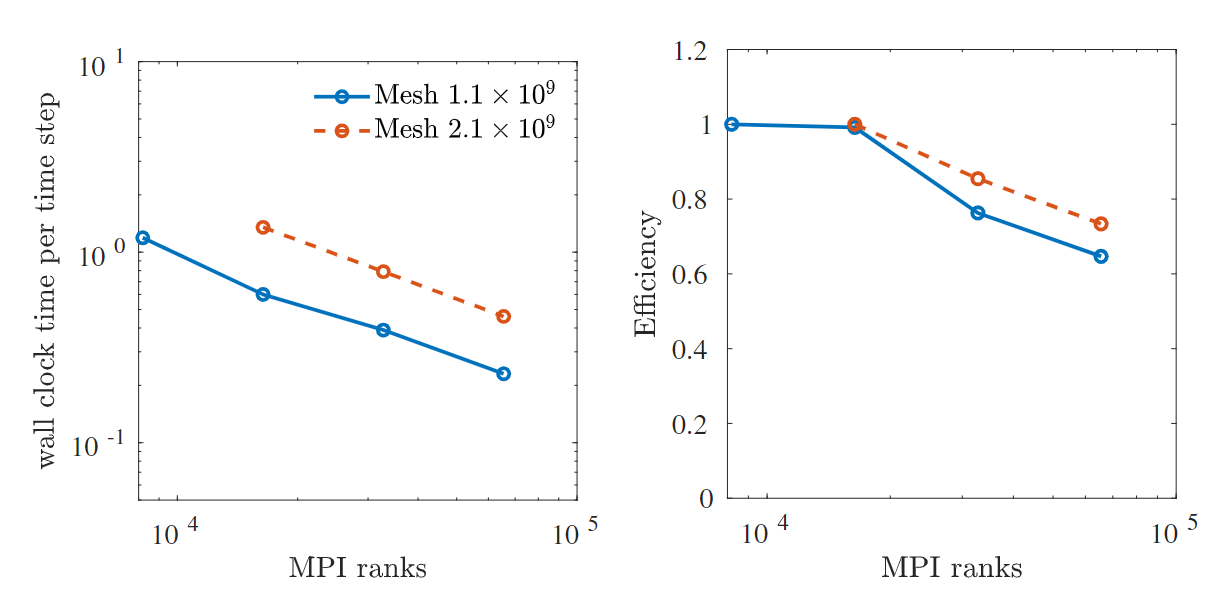}
\caption{Strong scaling behavior of the solver on ‘HAWK’, wall clock time per time step on the left side. On the right side the efficiency for different numbers of MPI ranks.}
\label{fig:9}
\end{figure}

The scalability of our direct numerical simulations, employing a high-order solver, was thoroughly evaluated on the advanced HPC system, {\it HAWK}, utilizing CPU cores ranging from 8,192 to 65,536. By leveraging ample computational resources, we achieved remarkable performance, with a wall clock time per time step ranging from 0.2 to 0.4 seconds. This optimal computational efficiency enables efficient direct numerical simulations of turbulent flows.
Notably, when utilizing the maximal number of MPI ranks, our simulations demonstrated a minimum efficiency of 70$\%$. This achievement highlights the exceptional scalability of our high-order solver on the state-of-the-art HPC cluster, {\it HAWK}, setting it apart from previous scalability tests conducted on the low-order Finite Volume Method (FVM) code OpenFOAM \cite{Evrim.2020, Chu.2016, Chu.2016c, Pandey.2017, Pandey.2018, Pandey.2018b, Yang.2020} using earlier HPC machines at HLRS.
Our findings underscore the outstanding scalability exhibited by our current high-order solver on the advanced HPC system, offering enhanced capabilities for conducting direct numerical simulations. This scalability is crucial for effectively studying turbulent flows and holds significant promise for advancing research in the field.

\section{Conclusions}
\label{sec:6}

Our study introduces a combined Convolutional Deep Autoencoder–Echo State Network (CDAE-ESN) model as a powerful tool for directly analyzing and forecasting the dynamics and low-order statistics of coupled turbulent channel-porous media flows. The significance of such a flow model lies in its applicability to various industrial systems, including transpiration cooling and smart interface engineering. We used fully-resolved direct numerical simulation to generate the data.
By integrating deep autoencoder and echo state network techniques, we overcome the challenges posed by the complex geometry of coupled flow systems, enabling effective modeling and prediction of dominant flow behaviors, particularly within the laminar regime of the porous domain. The obtained results demonstrate the capability of our proposed approach to capture and represent the intricate flow dynamics.
Furthermore, to broaden the model's applicability, we conducted fine-tuning on a dataset with varying porosity. This step aimed to enhance the generalizability of the CDAE-ESN model across a wider range of data domains. The reasonable agreement observed in the average statistics further validates the robustness and versatility of our approach.
In summary, the combined CDAE-ESN model presents a promising avenue for analyzing and predicting the behavior of coupled turbulent channel-porous media flows. Its potential applications in diverse industrial systems underscore its significance and highlight the future prospects for incorporating such data-driven models in fluid mechanics research.

\section{Acknowledgement}
The study has been financially supported by SimTech (EXC 2075/1–390740016) and SFB-1313 (Project No.327154368) from Deutsche Forschungsgemeinschaft (DFG). The financial support by the Deutsche Forschungsgemeinschaft with Grant No. SCHU 1410/30-1 and in parts by the project “DeepTurb – Deep Learning in and of Turbulence" by the Carl Zeiss Foundation is deeply appreciated. The authors gratefully appreciate the access to the high performance computing facility \emph{Hawk} at HLRS, Stuttgart of Germany. We would like to extend our sincere appreciation to Prof. Jörg Schumacher for his fruitful discussions.

\bibliography{reference}
\bibliographystyle{spbasic}

\end{document}